\documentclass[pra,aps,twocolumn,eqsecnum,showpacs]{revtex4}

\usepackage{dcolumn}
\usepackage{amsmath}
\usepackage{graphics}

\begin{document}

\title{Light trapping in high-density ultracold atomic gases for quantum memory applications}

\author{I.M. Sokolov$^{1,2}$ and D.V. Kupriyanov$^{1}$}%
\affiliation{$^{1}$Department of Theoretical Physics, State
Polytechnic University, 195251, St.-Petersburg, $^{ }$
\newline $^{2}$Institute for Analytical Instrumentation, Russian Academy
of Sciences, 198103, St.-Petersburg}
\email{IMS@IS12093.spb.edu}%

\author{R.G. Olave and M.D. Havey}
\affiliation{Department of Physics, Old Dominion University,
Norfolk, VA 23529}
\email{mhavey@odu.edu}

\date{\today }

\begin{abstract}
High-density and ultracold atomic gases have emerged as promising
media for storage of individual photons for quantum memory
applications. In this paper we provide an overview of our
theoretical and experimental efforts in this direction, with
particular attention paid to manipulation of light storage (a)
through complex recurrent optical scattering processes in very
high density gases (b) by an external control field in a
characteristic electromagnetically induced transparency
configuration.
\end{abstract}

\pacs{34.50.Rk, 34.80.Qb, 42.50.Ct, 03.67.Mn, 42.25.Dd, 42.50.-p}%

\maketitle%

\section{Introduction}
Manipulation of photonic degrees of freedom plays an important
role in many aspects of quantum information sciences
\cite{Nielsen}. One of these is the storage of photonic
information via a quantum memory, which forms an essential element
of an optical quantum repeater \cite{Briegel}. One candidate for
such a memory is an ultracold atomic gas \cite{Duan}, which can
have such desirable qualities as very narrow and thus selectable
resonance response, relative ease of manipulation of the atomic
density, and precise control of many aspects of the atomic
physical environment \cite{Metcalf,Grimm}. Long lived atomic
coherence may readily be generated, facilitating the formation, in
lower density gases, of dark state polaritonic excitations using,
for example, an optical $\Lambda$ configuration
\cite{Lukin1,Lukin2,Chaneliere,Matsukevich}. Various environmental
factors limit the lifetime of these coherences, and hence the time
scale of efficient optical storage in the medium. An additional
important factor that can limit the atomic coherence is multiple
light scattering of near resonance radiation in an optically thick
atomic ensemble \cite{Datsyuk1}. Multiple light scattering in
atomic gases is quasielastic, and so the scattered light can
remain in a coherent state.  This situation corresponds in the
lower density case to the so-called weak localization regime,
where such scattering can produce macroscopic observables such as
the coherent backscattering cone
\cite{Akkerman1,Labeyrie2,Bidel,Kupriyanov1,Balik,Labeyrie3,Labeyrie4,Muller,Antilocalization,Havey1,Kupriyanov2,Labeyrie1}.
The weak localization case is characterized by the inequality
$k\emph{l}$ $\gg$ $1$, where $k$ = $2\pi / \lambda$ is the light
wave vector, and $\emph{l}$ is the mean-free path for light
scattering.  Such an effect promises to help clarify the role of
multiple light scattering for quantum memory applications in this
case.  For much higher atomic densities, such that $k\emph{l}$
$\sim$ $1$, light scattering enters the so called strong
localization regime.  In this case, recurrent light scattering can
lead to formation of long lived atomic-photonic excitations  which
are currently under investigation as possible quantum  memories,
in searches for Anderson localization of light
\cite{Anderson,Sheng,Wiersma,Storzer,Storzer2}, development of
atomic-physics based random lasers \cite{Kaiser1,Perez}, and for
studies of single and multiple photon cooperative scattering
\cite{Scully1,Svidzinsky}.

In this paper we provide a brief overview of our theoretical and
experimental light storage programs using as approaches either
light localization or electromagnetically induced transparency.
This is accompanied by new results in several areas which
illustrate some of the factors important in the physical
processes. We concentrate particularly on the influence of
multiple scattering in ultracold clouds, and its role in the
timescale associated with quantum information storage.

\section{Light storage and light localization}
\subsection{Introduction}
We have recently reported a comprehensive description of our
theoretical approach to treatment of light scattering in very high
density and ultracold atomic gases \cite{Kupriyanov3}.  The atomic
density used in these calculations corresponds closely to the
Ioffe-Regel condition for light localization. In
\cite{Kupriyanov3} we used two different but complementary
approaches. In the first, we employed a self-consistent
description of the atomic sample in the spirit of the Debye-Mie
model for a macroscopic spherical scatterer consisting of a dense
configuration of atomic dipoles. In the second, we take a
microscopic approach in order to make exact numerical analysis of
the quantum-posed description of the single photon scattering
problem.

In the microscopic approach, the total scattering cross section
can be determined by the T-matrix, which in turn can be expressed
by the total Hamiltonian $H=H_{0}+V$of the joint atomic-field
system and by its interaction part $V$ as $T(E)=V+V(E-H)^{-1}V.$
In the rotating
wave approximation the internal resolvent operator $%
(E-H)^{-1}$ contributes to $T(E)$ only by being projected on the
states consisting of single atom excitation, distributed over the
ensemble, and the vacuum state for all the field modes. Defining
such a projector as P the projected resolvent
$\tilde{R}(E)=P(E-H)^{-1}P$ performs a finite matrix of size
determined by the number of atoms N and the structure of the
atomic levels. For the considered $F = 0\leftrightarrow F'=1$ transition $\tilde{R}%
(E)$ is a 3N x 3N matrix.

For a dipole-type interaction between atoms and field, the resolvent $\tilde{%
R}(E)$ can be found as the inverse matrix of the following operator $\tilde{R}%
^{-1}(E)=P(E-H_{0}-VQ(E-H_{0})^{-1}QV)P$, where the complementary projector $%
Q=1-P$, operating in the self-energy term, can generate only two
types of intermediate states: a single photon plus all the atoms
in the ground state; and a single photon plus two different atoms
in the excited state and others are in the ground level. For such
particular projections there is the following important constraint
on the interaction Hamiltonian: $PVP=QVQ=0$, which is apparently
valid for a dipole-type interaction $V=-\Sigma
\mathbf{d}_{j}\mathbf{E}_{j}$, where $\mathbf{d}_{j}$ is a dipole
operator of the j-th atom and $\mathbf{E}_{j}$ is the microscopic
displacement field at the point where the atom is located. Due to
this constraint the series for the inverse resolvent operator
$\tilde{R}^{-1}(E)$ is expressed by a finite number of terms and
explicit analytical expression for $\tilde{R}^{-1}(E)$ is
obtained. The resolvent $\tilde{R}(E)$ and T-matrix can be
calculated numerically for an atomic system consisting of a
several thousands of atoms. Thus, the microscopic approach gives
us the exact value of the scattering amplitude for macroscopic
atomic ensembles.

In this section, we present further elaboration of theoretical
results obtained in \cite{Kupriyanov3}, with particular emphasis
on the spectral response of the total scattering cross section,
and the time evolution of the total light scattered from a
spherical and high density sample, because such evolution is
strongly connected with light trapping in the cloud.  We also
present numerical results comparing the so-called vector and
scalar approaches to the microscopic scattering problem.  These
important results show clearly that the frequently-used scalar
model fails in a significant way for higher atomic densities
$\sim$ 5 $\cdot$ 10$^{13}$ atoms/cm$^{3}$.

In this work, our theoretical discussion of time dependent
scattering is based on the microscopic approach. The knowledge of
the resolvent operator $\tilde{R}(E)$ allows us to describe the
interaction of the atomic ensemble with weak coherent light, which
can be approximated as a superposition of the vacuum and single
excitation states. Thus, considering the light pulse as a
superposition of such coherent components, we generalize our
approach to the case of non steady state.

\subsection{Theoretical results}

\begin{figure}[htpb]
\includegraphics{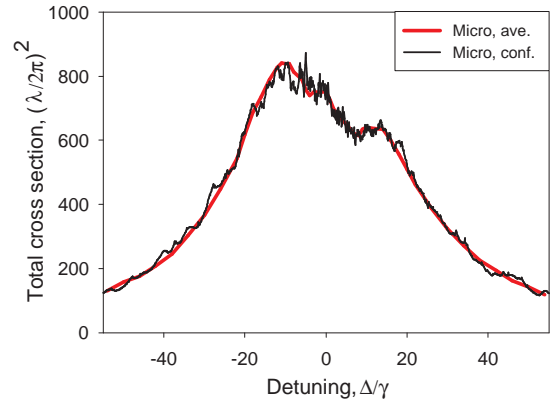}%
\caption{The spectral dependence of the total cross sections for
atomic samples of size $10\lambdabar$ and density
$n_0\lambdabar^{3}=0.5$. Calculations of the microscopic response
for a particular configuration (black curve) indicates a
micro-cavity structure generated by interatomic interactions in
dense disordered media.}
\label{fig1}%
\end{figure}

We first present in Fig. 1 the spectral variation of the total
light scattering cross section from a dense spherical cloud of
ultracold atoms.  For these calculations the sample radius is
$10\lambdabar$ and the atom density n = $n_0\lambdabar^{3}=0.5$.
As we mainly focus in our program on experiments with atomic Rb,
$\lambdabar$ = $780 / 2\pi$ nm, and the base unit of atomic
density (n = 1) corresponds to 5.2 $\cdot$ 10$^{14}$
atoms/cm$^{3}$. There are two results shown.  The first
corresponds to the scattering cross section for a single
realization of the atomic sample. This means that there is a
single configuration of atomic positions within the sample.  The
second result, indicated by the thicker curve (red on line), is
the average result obtained from many such configurations of
atomic positions.  We first see that, in both cases, the spectral
variations extend over a quite wide range, expressed in units of
$\gamma$. This is qualitatively due to the high optical depth
associated with transmission of light through the sample. Second,
we observe that both the single realization and the configuration
averaged cross sections contain broad oscillations which are
primarily due to diffractive scattering from the nearly opaque
spherical atomic sample.  The most important feature of these
results is the microstructure superimposed on the broad spectral
variations.  Of particular interest in the context of long lived
photonic modes within the sample is the existence of the very
narrow resonances apparent in the figure.  These features arise
from poles in the resolvent which have associated very narrow
spectral widths. We point out that there is a nearly continuous
spectral distribution of such poles, but that a significant
fraction of them have associated narrow widths for higher atomic
densities. In addition, we emphasize that the spectral locations
of these poles are configuration dependent; the resonance
locations are different for different spatial arrangements of the
atoms in the sample.

\begin{figure}[htpb]
\includegraphics{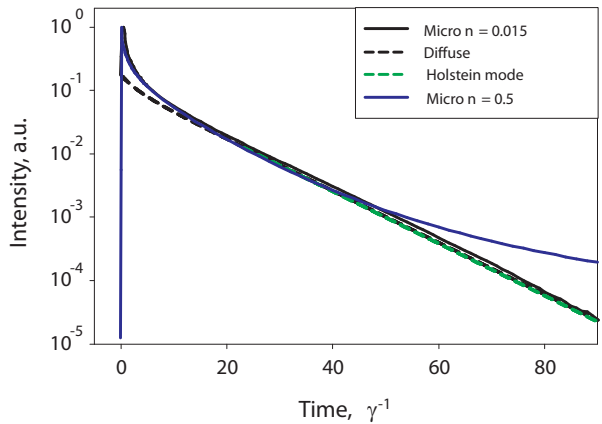}%
\caption{Comparison of the time dependence of the scattered light
intensity for spherical atomic clouds of different densities, but
approximately constant optical depth.  Preparation of the sample
is made with a temporally short pulse having a correspondingly
large enough bandwidth to cover the spectral response of the
system, viz. Fig. 1.  Microscopic results are compared with a
diffusive model, and with the expected Holstein mode decay for
each case.}
\label{fig2}%
\end{figure}

The microstructure apparent in Fig. 1 implies that there will be
long lived collective states present in any single configuration
of the atomic sample.  In the time domain, then, we can expect
evolution of light scattered from the sample to have associated
relatively slowly decaying components, measured in terms of the
characteristic time scale $1/ \gamma$. In Fig. 2 we show just such
behavior for the conditions associated with the curves of Fig. 1.
In the figure we first point out the characteristic exponential
decay of the scattered light normally expected at lower densities.
The three nearly identical curves correspond to a lower density
microscopic calculation (n = 0.015), to a  solution to the
diffusion equation for similar conditions, and to the decay
constant corresponding to the longest lived  Holstein mode.  At
the higher density associated with Fig. 1, we see a dramatic
slowing of the decay of the intensity at the longest times.  This
slowing occurs in coincidence with the development of the spectral
microstructure of Fig. 1, and corresponds to formation of
subradiant atom-field collective  modes in the sample.

\begin{figure}[htpb]
\includegraphics{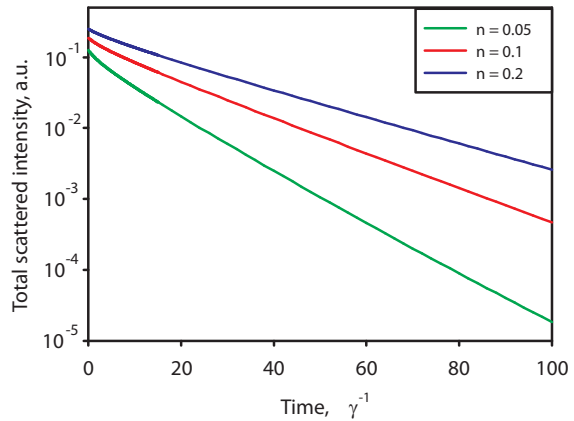}%
\caption{Time dependence of the total scattered light intensity
from a uniform spherical sample of atoms of radius $10\lambdabar$
and several different atom densities. Preparation of the sample is
made with a temporally short pulse having a correspondingly large
enough bandwidth to cover the spectral response of the system,
viz. Fig. 1.}
\label{fig3}%
\end{figure}

The time evolution of the scattered light intensity also shows
systematic variations with other experimentally accessible
variables.  For example, we show in Fig. 3 the longer time
intensity decay for a spherical sample of radius $10\lambdabar$
and several different atom densities.  In Fig. 3 it is apparent
that for higher densities the time scale for decay slows with
increasing density, this being a reflection of the combined
influence of the increasing diffusive decay time and the formation
at higher densities of microresonances as in Fig. 1. We emphasize
here that the time evolution of the afterglow of the excited
atomic sampled strongly depends on the conditions of excitation.
For instance, the results in Fig. 2 and Fig. 3 are obtained for
excitation of the cloud by a very short pulse, the spectrum of
which encompasses the entire manifold of excited states of the
ensemble.  Alternatively, we can excite the cloud with a
temporally longer pulse, one which has an associated narrower
spectrum.  This allows us to select spectral regions with the
highest density of longer-lived states.

The long time exponential decay for two different spectral
detunings from resonance are shows in Fig. 4.  These curves
correspond to excitation with a long, and thus spectrally narrow,
pulse, as discussed in the previous paragraph.  Shown are the
associated decay constants, from which we see that the rate of
decay is slower for larger detunings in this case. As discussed
previously, this connects directly with the spectral distribution
of collective states having different lifetimes; to be more
precise, with the spectral distribution of resolvent poles with
different imaginary parts.  The existence of long lived states is
determined by the strong interatomic interactions, which also
causes large spectral shifts of such states.  To effectively
generate excitation of these states with a long pulse, we must use
light with its central frequency essentially shifted from the free
atom resonance frequency.

\begin{figure}[htpb]
\includegraphics{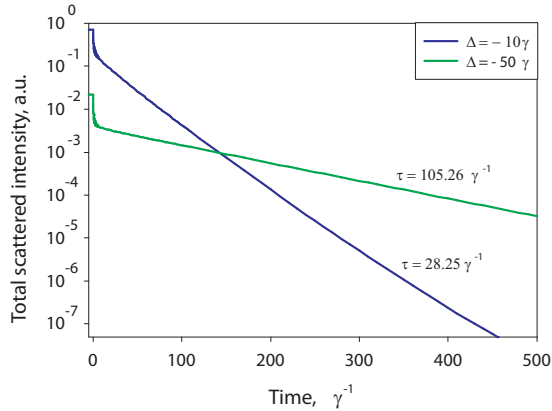}%
\caption{Time dependence of the total scattered light intensity
from a uniform spherical sample of atoms for two different
detunings from bare resonance. Excitation is with a temporally
long pulse, corresponding to a narrower pulse bandwidth than the
results of Fig. 2-3. The atom density n = 0.1.}
\label{fig4}%
\end{figure}

To close this section, we consider the results of Fig. 5.  These
calculations of the time evolution of the total light intensity
are made for a spherical sample of density n = 0.1 and show the
clear effects of using a more-correct vector model of the
light-matter microscopic interaction in comparison with a scalar
one.  Although we have shown such effects to be quite negligible
for lower atomic densities, for the higher density cases of
importance to light trapping, light localization and random
lasing, for example, the scalar approximation should not in
general be made in order to obtain the most reliable results.

\begin{figure}[htpb]
\includegraphics{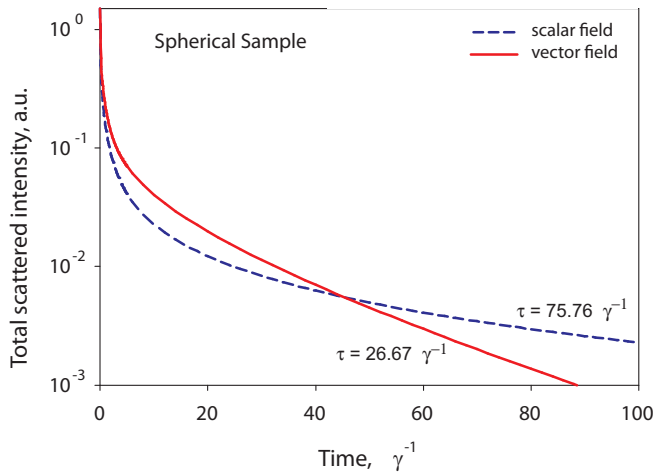}%
\caption{Comparison of vector and scalar field results for the
time evolution of the total scattered light intensity in a dense
medium: n = 0.1}
\label{fig5}%
\end{figure}

\section{Light storage and electromagnetically induced transparency}
\subsection{Introduction}
A second area of theoretical and experimental research interest in
our group is multiple light scattering in a medium dressed with a
strong control field.  In this section, we first present a
synopsis of our theoretical approach and some results taken from
Datsyuk, $\emph{et al.}$ \cite{Datsyuk1}.  This is followed by an
overview of more recently completed initial experiments in a
similar electronic $\Lambda$ configuration \cite{Olave}; these
results indicate that, along with the forward scattered beam, the
light scattered out of the coherent forward beam also propagates
under conditions of electromagnetically induced transparency. Here
we briefly summarize the theoretical and experimental approaches,
but we refer the reader to those papers for detailed exposition of
our theoretical and experimental studies, and focus here on some
illustrative results.

\subsection{Theoretical and experimental approaches and results}

\subsubsection{Theory}
Light transport in a dilute medium generally performs a
diffusion-type process, which in a semiclassical picture can be
visualized as a forwardly propagating wave randomly scattered by
atoms in a sample. In an optically dense medium this process
generates a zigzag-type path consisting of either macroscopically
or mesoscopically scaled segments of forwardly propagating waves.
To describe such a process theoretically we solve three problems.
First we determine the scattering tensor amplitude for an
arbitrary elementary incoherent scattering event. Then we describe
forward propagation of the light between two successive incoherent
scattering events. This is theoretically accomplished through the
Green's function formalism. This function is completely described
by the macroscopic susceptibility tensor of the atomic medium,
which in the considered case of a cloud of ultracold atoms under
EIT conditions is spatially inhomogeneous and optically
anisotropic (for more detail, see \cite{Datsyuk1}). In spite of
this inhomogeneity and anisotropy we are able to solve
analytically the system of Dyson-type equations for the
polarization components of the retarded Green function for the
light. Knowledge of the analytical expressions for both the
scattering tensor and photon Green function allows us to solve the
last, and third component of this problem, which is averaging over
all possible random scattering chains. This is done through the
framework of the procedure of Monte-Carlo simulation. Realization
of this theoretical approach \cite{Datsyuk1} allows us to describe
the polarization, the spectral and the temporal properties of the
scattered light.

\begin{figure}[htpb]
\includegraphics{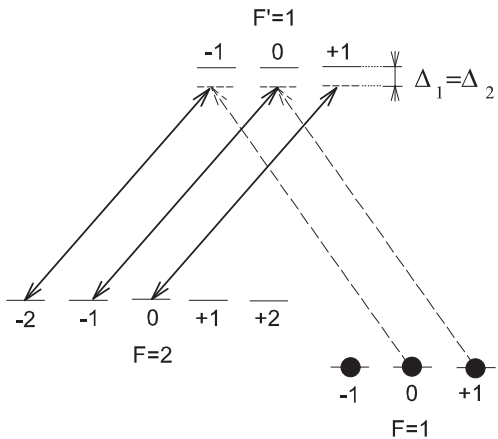}%
\caption{An example of an excitation scheme for observation of the
EIT effect in the system of hyperfine and Zeeman sublevels of the
$D_1$-line of ${}^{87}$Rb. The coupling field is applied with
right-handed circular polarization to the $F=2\to F'=1$ transition
and the probe mode in the orthogonal left-handed polarization
excites the atoms on the $F=1\to F'=1$ transition. The EIT effect
appears for equal detunings of the coupling and probe modes from
atomic resonances: $\Delta_1=\Delta_2$}
\label{fig6}%
\end{figure}

One of the more important results from our earlier theoretical
papers \cite{Datsyuk1} is that the light scattered out of the
coherent probe beam continues to propagate through multiple
scattering in the medium dressed by the external control field.
This means that the role of this light in the subsequent storage
and retrieval of any photonic pulse has to be treated properly;
that is, the coherence associated with the multiple scattering
process has to be taken into account.

We have used the scheme of Fig. 6 to illustrate features of this
effect.  In that scheme, which refers to the $D_1$-line of
${}^{87}$Rb, the coupling field is applied with right-handed
circular polarization to the $F=2\to F'=1$ transition. The probe
mode, on the other hand, is applied in the orthogonal left-handed
polarization and excites the atoms via the $F=1\to F'=1$
transition. The two fields are assumed to be in two-photon
resonance. The temporal profile of the probe intensity is taken to
be Gaussian with a width of 100 $\gamma^{-1}$.  The pulse reaches
its peak intensity at a time t = 0.  Examination of the results
shown in Fig. 7 shows that the light emerging from the sample has
a significant extension to longer times, particularly for the
Rayleigh $\sigma_{+} \rightarrow h_{-}$ scattering channel.  This
is a manifestation of the well known slow-light effect in the
scattering channels.  This effect is normally associated with the
forward scattering of the probe beam, but the results of
\cite{Datsyuk1} clearly show that the sample dressing by the
control field plays a significant role in the scattered light
dynamics.  Finally, we point out that the temporal beating in the
scattered light intensity is a result of the spectral distortion
of the incident pulse by the EIT prepared medium.  In other words,
the Gaussian spectral distribution associated with the incident
pulse is transformed by the medium, and particularly by the
transparency associated with the EIT effect. The spectral hole
generated in the probe spectrum results in beating at the
frequency associated approximately with the width of the EIT
transparency.

\begin{figure}[htpb]
\includegraphics{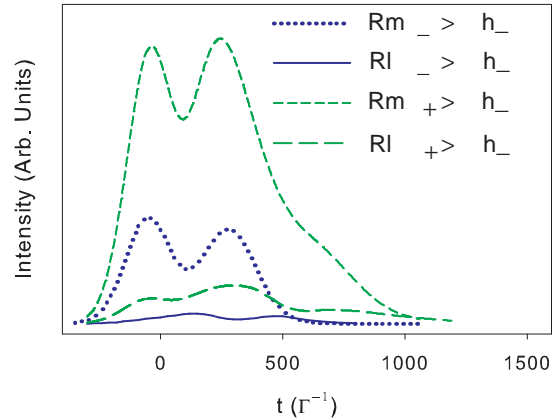}%
\caption{The intensity profiles for the portion of the light pulse
scattered at $90^0$ to the direction of the incident pulse. The
curves represent the Rayleigh (Rl) channel ($F=1\to F'=1\to F=1$)
and the Raman (Rm) channel ($F=1\to F'=1\to F=2$), with the input
polarization state, and polarization channel of the emerging light
as indicated in the caption. In all cases, the observation channel
corresponds to detection of light with left-hand helicity
($h_{-}$).}
\label{fig7}%
\end{figure}

\subsubsection{Experiment}

\begin{figure}[htpb]
\includegraphics{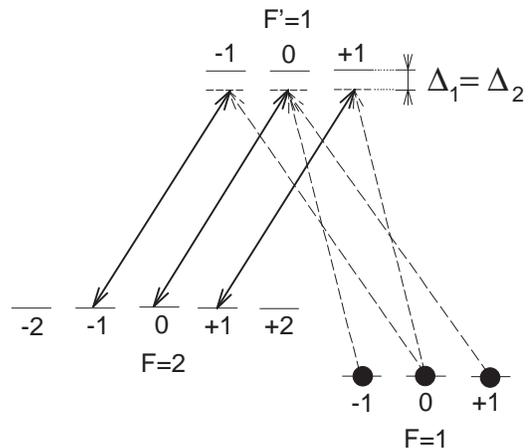}%
\caption{Experimental probe and control field excitation
configuration associated with the $F=1\to F'=1\to F=2$ transition
of the $D_2$ line in $^{87}$Rb. The control field is indicated by
the solid lines, and the probe by dashed lines. The illustration
corresponds to two photon resonance; Zeeman level light shifts
$\sim$ 1 MHz associated with off resonance transitions are not
shown.}
\label{fig8}%
\end{figure}

The experimental program focuses on ultracold atomic samples
prepared and confined in a magneto optical trap (MOT). We present
here highlights of the experimental instrumentation: details are
presented elsewhere \cite{Olave}. The $^{87}$Rb MOT is arranged in
a standard vapor loaded configuration, and consists of samples of
about $10^{7}$ atoms with a Gaussian radius of 0.3 mm. The sample
has an optical depth, on the $F=2\to F'=3$ trapping transition, of
about 10, while the atom sample has a typical temperature $\sim$
100 $\mu$K.  These samples are of sufficient optical depth that
there can be several orders of multiple scattering of light
scattered out of the probe beam. The excitation configuration we
have used in our initial experiments to study the dynamics of the
scattered light is shown in Fig. 8. In this scheme, the linearly
polarized control field (z quantization axis) is tuned in the
vicinity of the $F = 2 \rightarrow F' = 1$ transition, while the
much weaker linearly polarized (x direction) probe beam is tuned
near the $F = 1 \rightarrow F' = 1$ transition. The two lasers are
characteristically tuned to be near two photon resonance, where
$\Delta_1 = \Delta_2$.  We point out that this configuration is
not optimal for obtaining large orders of multiple scattering. The
main reason for this is that there are open decay channels to the
F = 2, m = $\pm$ 2 states, in which population can be trapped for
relatively long periods of time.

\begin{figure}[htpb]
\includegraphics{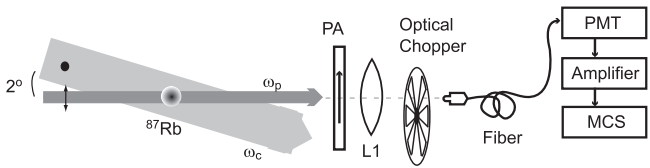}%
\caption{Schematic diagram of the time evolution of transmitted,
stored, and retrieved using the level scheme of Fig. 8.  PA stands
for polarization analyzer, PMT represents a photomultiplier tube,
and MCS indicates a multichannel scalar.}
\label{fig9}%
\end{figure}

The basic experimental approach is shown schematically in Fig. 9.
In the figure, it is seen that the pump and control fields,
depicted by $\omega_p$ and $\omega_c$, respectively, propagate in
approximately collinear fashion through the sample. The lasers are
external cavity diode lasers, locked to saturated absorption
resonances in a Rb vapor cell, and frequency tuned and switched in
the conventional way with acousto optical modulators.  Each laser
has a long time average bandwidth on the order of 500 KHz. In
these initial experiments, the two lasers are not mutually phase
locked; the combined laser bandwidths then provide the ultimate
limit to the characteristic ground state coherence lifetime. In
the experiment, a small angle is arranged between them in order to
suppress background due to the much more intense control field on
top of the relatively weak probe beam intensity. Further
suppression is achieved by means of a linear polarization analyzer
(PA) placed directly in the path of the probe and control beams.
An optical chopper is used to minimize intense MOT fluorescence
signals while the sample is being loaded; the chopper trigger is
used as a master switch to initiate each data cycle, minimizing
the relative absolute instability of the chopper timing. The probe
laser beam is launched into an optical fiber which transmits the
light to a photomultiplier operating in a photon counting mode.
Following amplification of the individual pulses, the signals are
time sorted with up to 5 ns resolution in a multichannel scalar
(MCS).  The MCS also serves to accumulate and store the data for
later analysis.

Under conditions of two photon resonance, and for typical control
and probe field intensities, we have found that, in our
experimental conditions, it takes about 500 ns for the transmitted
light to reach an approximately steady state level. With this
configuration, we have been able to observe slowed transmission of
the forward scattered probe beam, and also to measure so-called
stopped light pulses for a range of different delay times.  In
these measurements, which we present elsewhere \cite{Olave}, the
control and probe fields are turned off in an approximately
adiabatic way; later reapplication of the control field to the
sample generates, from the dark state polariton, a retrieval of
the forward scattered pulse. The lifetime of the regenerated
pulses is quite short in the present experiment, and is limited by
the mutual coherence of the ground state hyperfine levels.  This
in turn is determined by the relative phase instability of the
control and probe fields. It is the storage and retrieval of such
light pulses that provides an intriguing option for using EIT as
the basis for an optical memory.

However, the primary focus of our projects is comparison of the
forward scattered light with the sideways, or diffusely scattered
light.  To illustrate that comparison, we present in Fig. 10 the
transparency in the forward scattered light as a function detuning
of the weak probe beam from two photon resonance. These data were
taken for a control field Rabi frequency of 1.2 $\gamma$.  From
the figure, we see that there is a limited transparency in the
transmitted light due to the relative sizes of the control Rabi
frequency and the bandwidth due to the quite short ground state
coherence time; larger control Rabi frequencies lead to nearly
complete transparency \cite{Olave}.  The corresponding suppression
of diffusely scattered light in the vicinity of two photon
resonance is evident in the upper panel of Fig. 10. Theoretical
analysis of the time evolution of this scattered light suggests
that it propagates with a reduced group velocity, just as the
forward scattered light.  Our current projects are focused towards
establishing the coherence properties of the multiply scattered
light through observation of the coherent backscattering cone
\cite{Datsyuk1}.

Finally, we point out that the present arrangement is not optimum,
both because of the configuration used and because of the limited
scale of the ground state coherence; further research is underway
to extend this scale, and the amount of multiple scattering, in
order to explore the role of coherent multiple scattering on the
ultimate lifetime of light storage in ultracold and dense atomic
gases. Further details of these experiments, including the effects
of the control dressing field on the propagation of the light
scattered from the probe pulse, are presented elsewhere
\cite{Olave}.

\begin{figure}[htpb]
\includegraphics{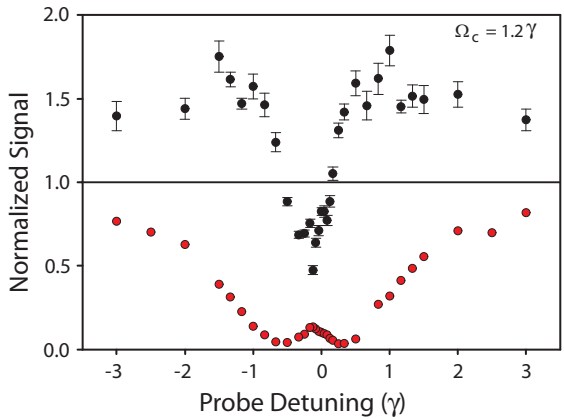}%
\caption{Comparison of forward and diffusely scattered light using
the level configuration of Fig. 8, and the experimental scheme of
Fig. 9. The lower panel (red circles) refers to transmission,
while the upper panel (black circles) refers to the diffusely
scattered data. The data correspond variations of the weak probe
laser frequency around two photon resonance, and to a control
field Rabi frequency $\Omega_c$ = 1.2 $\gamma$; with reference to
Fig. 8, $\Omega_c$ refers to the average control Rabi frequency
for the three indicated transitions.}
\label{fig10}%
\end{figure}

\section{Summary}
In this paper we have briefly summarized some aspects of our
theoretical and experimental studies of light storage in dense and
ultracold atomic gases. The two approaches we use have as a common
theme the role of coherent multiple scattering in such media.  For
lower atomic densities, multiple scattering may serve as a
limiting factor in photon storage time.  There also is the
intriguing possibility that a new type of quasiparticle, a diffuse
dark state polariton, may be formed.  Experiments in that
direction are underway.  For greater atomic densities, recurrent
multiple scattering may lead to formation of long lived subradiant
modes; these modes may serve in themselves as a means to store
photonic information.  Control of these modes may be achieved by
manipulating dynamically the optical depth of the atomic media.
One way to achieve this is through rapid control of the light
shift of the involved atomic resonance levels.

\section{Acknowledgments}
We appreciate the financial support of the Russian Foundation for
Basic Research (Grant No. RFBR-08-02-91355), the National Science
Foundation (Grant No. NSF-PHY-0654226).

\end{document}